\documentclass[12pt,a4paper]{article}
\usepackage{pdflscape}
\usepackage{amsmath,amssymb,amsthm,amsxtra,overpic,bbm,bm,epsfig,subfigure}
\usepackage{mathrsfs}
\usepackage{booktabs}
\usepackage{graphicx}
\usepackage{xcolor}
\usepackage{comment}
\usepackage{epstopdf}
\usepackage{float}
\usepackage{cite}
\usepackage{array}
\usepackage{makecell}
\usepackage{enumitem}
\usepackage{tabularx} % for 'tabularx' environment
\usepackage{overpic}
\usepackage{fancyhdr}
\usepackage{float}
\usepackage{rotating}
\usepackage{color}
\makeatletter
\usepackage{slashed,stmaryrd}
\def\thefootnote{\fnsymbol{footnote}}
\usepackage{geometry}
\definecolor{MyDarkBlue}{rgb}{0.1, 0.1, 0.8} %defining the color 'MyDarkBlue'
\definecolor{SBlue}{rgb}{0.2, 0.4, 0.7} %defining the color 'MyDarkBlue'
\definecolor{MyLightBlue}{rgb}{0.22,0.51,0.9}
\definecolor{MyGreen}{rgb}{0.0, 0.5, 0.0}
\definecolor{BrickRed}{rgb}{0.8, 0.25, 0.33}
\usepackage{hyperref}
\hypersetup{colorlinks, citecolor=SBlue,linkcolor=MyDarkBlue, urlcolor=BrickRed}

\begin{document}
\begin{center}
{\Large \bf 
	Tritium beta decay with modified neutrino dispersion relations: KATRIN in the dark sea
}
\end{center}
\renewcommand{\thefootnote}{\fnsymbol{footnote}}
\vspace{0.05in}
\begin{center}
{
	{}~\textbf{Guo-yuan Huang$^{1}$}\footnote{ E-mail: \textcolor{MyDarkBlue}{guoyuan.huang@mpi-hd.mpg.de}} and
	{}~\textbf{Werner Rodejohann$^{1}$}\footnote{ E-mail: \textcolor{MyDarkBlue}{werner.rodejohann@mpi-hd.mpg.de}}
}
\vspace{0.1cm}
{
	\\
	\em $^1$Max-Planck-Institut f{\"u}r Kernphysik, Saupfercheckweg 1, 69117 Heidelberg, Germany
} 
\end{center}
\renewcommand{\thefootnote}{\arabic{footnote}}
\setcounter{footnote}{0}
\thispagestyle{empty}
\vspace{0.5cm}
%%%%%%%%%%%%%%%%%%%%%%%%%%%%%%%%%%%%%%%%%%%%%%%
%%%%%%%%%%%%%%%%%%%%%%%%%%%%%%%%%%%%%%%%%%%%%%%
\begin{abstract}
\noindent
We explore beta decays in a dark background field, which could be formed by  dark matter, dark energy or a fifth force potential. In such scenarios, the neutrino's dispersion relation will be modified by its collective interaction with the dark field, which can have interesting consequences in experiments using tritium beta decays to determine the absolute neutrino mass.
Among the most general interaction forms, the (pseudo)scalar and (axial-)vector ones are found to have interesting effects on the spectrum of beta decays.
In particular, the vector and axial-vector potentials can induce distinct signatures by shifting the overall electron energy scale, possibly beyond the usually defined endpoint. The scalar and pseudoscalar potentials are able to mimic a neutrino mass beyond the cosmological bounds.
We have placed stringent constraints on the dark potentials based on the available experimental data of KATRIN. 
The sensitivity of future KATRIN runs is also discussed. 
%For the $0\nu\beta\beta$ decays, the effect is significant only when the lepton-number-violating interaction with scalars is present, in which case the transition driven by the dark matter field is not suppressed by the small neutrino mass.
\end{abstract}
%%%%%%%%%%%%%%%%%%%%%%%%%%%%%%%%%%%%%%%%%%%%%%%
%%%%%%%%%%%%%%%%%%%%%%%%%%%%%%%%%%%%%%%%%%%%%%%
\setcounter{footnote}{0}

\newpage
%%%%%%%%%%%%%%%%%%%%%%%%%%%%%%%%%%%%%%%%%%%%%%%
%%%%%%%%%%%%%%%%%%%%%%%%%%%%%%%%%%%%%%%%%%%%%%%

\section{Introduction}
\noindent
The Mikheyev-Smirnov-Wolfenstein (MSW) effect of neutrino oscillations has been well established and understood~\cite{Wolfenstein:1977ue,Wolfenstein:1979ni,Mikheyev:1985zog,Mikheev:1986wj,Halprin:1986pn,Chang:1988yn,Kuo:1989qe,Sawyer:1990tw,Raffelt:1996wa,Cardall:1999bz,Fujii:2002mz,Linder:2005fc,Akhmedov:2012mk,Blennow:2013rca,Akhmedov:2020vua}.
The weak interactions between neutrinos and other fermions lead to a modified dispersion relation when neutrinos traverse matter.
Besides this established phenomenon, there is an interesting possibility that neutrinos feel a background of particles in the ``dark sea''. 
Both massive neutrinos and dark matter (DM) call for the extension of the Standard Model (SM), and it remains a mystery as to how they interact. 
Going beyond the weak interaction, secret neutrino interactions have been proposed as a helpful option to alleviate several observational tensions, e.g., the Hubble crisis~\cite{DiValentino:2021izs,Arias-Aragon:2020qip,Grohs:2020xxd,Blinov:2019gcj,Escudero:2019gvw,Forastieri:2019cuf,Huang:2021dba}, as well as eV and keV sterile neutrino hypotheses~\cite{Hannestad:2013ana,DeGouvea:2019wpf,Berryman:2022hds}. As for DM, the small scale structures are actually in favor of DM self-interactions (see, e.g., Refs.~\cite{Bullock:2017xww,Tulin:2017ara} for recent reviews).
In this regard, it is natural to expect that those two elusive sectors connect via some dark forces~\cite{Ma:2006km,Mangano:2006mp,Serra:2009uu,vandenAarssen:2012vpm,Wilkinson:2013kia,Wilkinson:2014ksa,Bertoni:2014mva,DiValentino:2017oaw,Olivares-DelCampo:2017feq,Escudero:2018thh,Hooper:2021rjc,Davoudiasl:2018hjw,Orlofsky:2021mmy,Coy:2022cpt,Kelly:2020pcy}. The $\nu$-DM interaction has already been proposed as a possible way to alleviate tensions between small-scale structure observations and $N$-body simulations~\cite{vandenAarssen:2012vpm}, even though it is constrained by the Big Bang nucleosynthesis~\cite{Ahlgren:2013wba,Huang:2017egl}. Moreover, a latest analysis of Lyman-$\alpha$ data has found hints for a strong $\nu$-DM interaction~\cite{Hooper:2021rjc}.

There might be various scenarios in which neutrinos feel a background of particles in the ``dark sea'', which can be formed by fifth force, ultralight DM, as well as dark energy (DE) field. By collectively interacting with the background fields, the neutrino kinematics is expected to be altered similar to the established MSW effect.
%~\cite{Berlin:2016woy,Brdar:2017kbt,Krnjaic:2017zlz,Liao:2018byh,Capozzi:2018bps,Reynoso:2016hjr,Huang:2018cwo,Pandey:2018wvh,Farzan:2018pnk,Choi:2019ixb,Baek:2019wdn,Choi:2019zxy,Choi:2020ydp,Dev:2020kgz,Baek:2020ovw,Losada:2021bxx,Smirnov:2021zgn,Alonso-Alvarez:2021pgy,Ge:2019tdi,Davoudiasl:2018hjw,Joshipura:2003jh,Grifols:2003gy,Grifols:2003gy,Bandyopadhyay:2006uh,Heeck:2010pg,Wise:2018rnb,Ge:2018uhz,Smirnov:2019cae,Babu:2019iml,Bustamante:2018mzu,Coloma:2020gfv}: 
In this context, neutrino oscillations in the dark potentials have been widely discussed in the literature~\cite{Berlin:2016woy,Brdar:2017kbt,Krnjaic:2017zlz,Liao:2018byh,Capozzi:2018bps,Reynoso:2016hjr,Huang:2018cwo,Pandey:2018wvh,Farzan:2018pnk,Choi:2019ixb,Baek:2019wdn,Choi:2019zxy,Farzan:2019yvo,Choi:2020ydp,Dev:2020kgz,Baek:2020ovw,Losada:2021bxx,Smirnov:2021zgn,Alonso-Alvarez:2021pgy,Ge:2019tdi,Davoudiasl:2018hjw,Joshipura:2003jh,Grifols:2003gy,Grifols:2003gy,Bandyopadhyay:2006uh,Heeck:2010pg,Wise:2018rnb,Ge:2018uhz,Smirnov:2019cae,Babu:2019iml,Bustamante:2018mzu,Coloma:2020gfv}. 
However, 
the neutrino oscillation experiments are only sensitive to relative values of the potential.
If three generations of neutrinos couple identically to the dark field, similar to the neutral $Z$-exchange in the Standard Model, the neutrino oscillation will be blind to the dark potential.
Furthermore, if ultralight dark matter is responsible for the dark potential,
and the dark field is fast oscillating over the neutrino baseline, neutrinos in flight will experience a vanishing averaged effect.
In this regard, the experiments probing absolute neutrino masses such as $\beta$ decays can be a better place to search for such dark potentials.
In this work, we perform a systematic analysis of the impact of dark potentials on $\beta$-decay neutrino mass searches in which kinematics is used to determine $m_\nu$. The emphasis will be put on the ongoing KATRIN experiment~\cite{KATRIN:2019yun,KATRIN:2021fgc,Aker:2021gma}, which has recently set the world-leading model-independent constraint on the absolute neutrino mass. However, the framework presented in this work should also in principle apply to other promising neutrino mass experiments, such as Project 8~\cite{Project8:2022wqh,Project8:2017nal,Project8:2014ivu,Project8:2022hun}, ECHo~\cite{Gastaldo:2017edk} as well as PTOLEMY~\cite{PTOLEMY:2019hkd}.
It is worthwhile to mention that KATRIN data have already been used to set constraints on the Lorentz violation parameter by searching for possible periodic signals~\cite{KATRIN:2022qou}.
Despite various realizations of the dark potential, our results of analysis will be presented in a form as model-independent as possible, such that they can be applied to various scenarios of model realizations.

The rest of this work is organized as follows. In Sec.~2, starting from general interaction forms we derive the modified equation of motion and dispersion relation of neutrinos. 
In Sec.~3, we give some examples of specific model realizations for completion.
In Sec.~4, we compute the beta-decay rate in the dark potential. In Sec.~5, we explore the experimental effect at KATRIN.
We make our conclusion in Sec.~6.

%\begin{eqnarray}\label{eq:L}
%-\mathcal{L} \supset
% \begin{dcases} 
%g_{\phi}^{} \, \phi \, \overline{\nu^{}_{}}  \nu^{}_{} \, \delta^{}_{\rm M} \; ,  \\[0.5ex]
%g_{\varphi}^{} \, \varphi \, \overline{\nu^{}_{}} \gamma^{}_{5}  \nu^{}_{} \, \delta^{}_{\rm M} \; ,  \\[0.5ex]
%g_{V}^{} V^{}_{\mu} \, \overline{\nu^{}_{}} \gamma^{\mu}  \nu^{}_{} \, \delta^{}_{\rm M}  \; , \\[0.5ex]
%g_{a}^{} a^{}_{\mu} \, \overline{\nu^{}_{}} \gamma^{}_{5}\gamma^{\mu}  \nu^{}_{} \, \delta^{}_{\rm M}  \; , \\[0.5ex]
%g_{T}^{} T^{}_{\mu \nu} \, \overline{\nu^{}_{}} \sigma^{\mu \nu} \nu^{}_{} \, \delta^{}_{\rm M}  \; . 
%\end{dcases}
%\end{eqnarray}
\section{\label{sec:2}Dark MSW effect}
Regardless of the nature of the underlying interaction, neutrinos will feel a ``dark potential'' that can have five different forms~\cite{Rosen:1982pj,Rodejohann:2017vup}: 
\begin{eqnarray}\label{eq:L}
-\mathcal{L} & \supset &
\left( g_{\phi}^{} \, \phi \, \overline{\nu^{}_{}}  \nu^{}_{} \,   +
g_{\varphi}^{} \, \varphi \, \overline{\nu^{}_{}} \gamma^{}_{5}  \nu^{}_{} + 
g_{V}^{} V^{}_{\mu} \, \overline{\nu^{}_{}} \gamma^{\mu}  \nu^{}_{}  + g_{a}^{} a^{}_{\mu} \, \overline{\nu^{}_{}} \gamma^{}_{5}\gamma^{\mu}  \nu^{}_{} +
g_{T}^{} T^{}_{\mu \nu} \, \overline{\nu^{}_{}} \sigma^{\mu \nu} \nu^{}_{} \right) \delta^{}_{\rm M}  \; . 
\end{eqnarray}
Here, $\delta^{}_{\rm M} = 1/2$ ($1$) if neutrinos are Majorana (Dirac) particles, and $\phi$,  $\varphi$, $V^{}_{\mu}$, $a^{}_{\mu}$ and $T^{}_{\mu \nu}$ represent real scalar, pseudoscalar, vector, axial-vector and tensor fields, respectively. 
The neutrino field is composed of $\nu = \nu^{}_{\rm L} + \nu^{\rm c}_{\rm L}$ for Majorana neutrinos, and $\nu = \nu^{}_{\rm L} + \nu^{}_{\rm R}$ for Dirac neutrinos. 
For Dirac neutrinos, the coupling constants $g^{}_{\phi}$, $g^{}_{V}$ and $g^{}_{a}$ are Hermitian matrices in general, while $g^{}_{\varphi}$ is anti-Hermitian. For the Majorana case, $g^{}_{\phi}$ and $g^{}_{a}$ are real symmetric matrices, while $g^{}_{\varphi}$ and $g^{}_{V}$ are purely imaginary symmetric and antisymmetric matrices, respectively.
The fields $\phi$,  $\varphi$, $V^{}_{\mu}$, $a^{}_{\mu}$ and $T^{}_{\mu \nu}$ do not necessarily correspond to real particles, e.g., they may represent  contributions from  coherent forward scattering or virtual fifth forces. Note that here we assume the field $X = \phi,  \varphi, V^{}_{\mu}, a^{}_{\mu}, T^{}_{\mu \nu}$ to be real, and only the overall sign combined with ``charge'' $g^{}_{X} X$ matters for neutrinos.  

A well-known example is already given by the standard MSW effect.
For instance, via the charged-current interaction with electrons in the Earth, the neutrino will feel a tiny potential of the form $g^{}_{V} V^{}_{0} + g^{}_{a} a^{}_{0} =  \sqrt{2} G^{}_{\rm F} n^{}_{e}$, where $G^{}_{\rm F}$ is the Fermi constant, $n^{}_{e}$ is the electron number density, and the isotropic spatial components $g^{}_{a}\bm{a}$ are averaged out.
Our results are simplified by imposing the assumption that the dark field is purely time-like, i.e., we assume from the cosmological principle that the preference of spatial orientation of the background is not significant or simply averaged out.
Hence, the antisymmetric tensor field $T^{}_{\mu\nu}$~\cite{Capozzi:2018bps} is  vanishing in our context.
%\footnote{This approximation is validated in the context under our discussion, as the motion of the solar system relative to the cosmic frame set by cosmic microwave background (dark matter as well) is ultra non-relativistic.}. 

%For the Majorana neutrino, the scalar interaction motivated by the Majoron model, violates the lepton number, or equivalently $\phi$ carries a lepton number of two. The neutrino-vector interaction could be generated if neutrinos are associated with some broken $U(1)$ hidden sector.

%An analogy of the tensor form is the interaction between the electromagnetic wave and the neutrino dipole moment, but here the tensor field may be generated from some dark photon background.
%& &(\partial^2 + m^2_{\phi})\phi = - g^{}_{\phi} \overline{\nu}  \nu \, \delta^{}_{\rm M}  , \label{eq:EOMphi} \\ 
%& &(\partial^2 + m^2_{V})V^{}_{\mu} = g^{}_{V}\overline{\nu} \gamma^{}_{\mu}  \nu \, \delta^{}_{\rm M}  ,  \label{eq:EOMV} \\ 

The equation of motion (EOM) of the neutrino wave function given the interactions in Eq.~(\ref{eq:L}) is described by
\begin{eqnarray}
& &\left[ \vphantom{\widehat{M}^{}_{\nu}}\gamma^{\mu }(i {\partial}^{}_{\mu} - g^{}_{V} V^{}_{\mu} + g^{}_{a} a^{}_{\mu}  \gamma^{}_{5} )  - (\widehat{M}^{}_{\nu} + g^{}_{\phi} \phi  + g^{}_{\varphi} \varphi \gamma^{}_{5}) \right] \nu = 0 \; ,\label{eq:EOMnu}
\end{eqnarray}
where the neutrino is written in the mass eigenstate basis in vacuum, therefore the mass matrix $\widehat{M}^{}_{\nu}$ is diagonal.
We can assume that three generations of neutrinos share the same coupling, i.e., $g^{}_{X} \propto \mathbbm{1} $ for all coupling constants, which in particular implies that there is no effect in neutrino oscillations. 
Note that for Majorana neutrinos, the diagonal vector interactions will be vanishing.
%Note that the above EOM of the neutrino applies to the general cases when the dark background is present irrespective of its origin.

The collective effect of the dark sea is to modify the dispersion relation of neutrinos, which can be obtained by multiplying $ \gamma^{\mu }(i {\partial}^{}_{\mu} - g^{}_{V} V^{}_{\mu} + g^{}_{a} a^{}_{\mu} \gamma^{}_{5} )  + (\widehat{M}^{}_{\nu} + g^{}_{\phi} \phi - g^{}_{\varphi} \varphi \gamma^{}_{5})$ to the left of Eq.~(\ref{eq:EOMnu}). 
For the plane-wave solution
we have $i \partial_{\mu}{\nu} =  p^{}_{\mu} {\nu}$, such that ${\nu} \propto \mathrm{exp}(-i\, E^{}_{\nu} \cdot t + i\, \bm{p}^{}_{\nu} \cdot \bm{x})$, where $E^{}_{\nu}$ is yet to be fixed.
%We do not require that the solution should belong to the positive- or negative-frequency mode so far, which means that energy and momentum here are still generic real numbers that can be negative.
Ignoring the cross terms of the background fields (i.e., we turn on only one field at a time), we end up with the following dispersion relation for neutrinos:
\begin{eqnarray}\label{eq:dispersion}
(E^{}_{\nu} - g^{}_{V} V^{}_{0} )^2 = \left(\left|\bm{p}^{}_{\nu}\right| - \hat{\bm{p}}\cdot \bm{\Sigma}  \,   g^{}_{a} a^{}_{0} \right)^2 + (\widehat{M}^{}_{\nu} +g^{}_{\phi} \phi )^2 + (|g^{}_{\varphi}|\varphi)^2   ,
\end{eqnarray}
where $E^{}_{\nu}$ and $\bm{p}^{}_{\nu}$ are the energy and momentum of the neutrino, $\hat{\bm{p}}$ denotes the direction of the momentum, and $\bm{\Sigma} \equiv \gamma^{}_{5}\gamma^{}_{0}\bm{\gamma}$ stands for the spin operator.
Eq.~(\ref{eq:dispersion}) has typically two energy solutions: the upper one (positive for most cases) corresponds to the particle $\nu^{(+)}$, and the lower one (mostly negative), which is not bounded from below, should be interpreted as the antiparticle $\nu^{(-)}$. For the antineutrino, the direction of the momentum $\hat{\bm{p}}$ should be reversed along with the energy accordingly.
%Note that the above dispersion relation applies to both the relativistic and non-relativistic cases. 
%The last term $+(|g^{}_{\varphi}|\varphi)^2 $ ($g^{}_{\varphi}$ is imaginary) contributed by the pseudoscalar interaction has a positive sign, different from the tachyon scenario.

Eq.\ (\ref{eq:dispersion}) implies that the scalar and pseudoscalar potentials $g^{}_{\phi} \phi$ and $|g^{}_{\varphi}|\varphi$ simply add to the vacuum mass term.
The vector potential $g^{}_{V} V^{}_{0}$ shifts the overall energy of neutrinos.
The axial-vector potential $ g^{}_{a} a^{}_{0}$ will lead to  helicity-dependent energies of the states with $\hat{\bm{p}}\cdot \bm{\Sigma}\, \nu^{(+)} = \pm \nu^{(+)}$ for neutrino and $-\hat{\bm{p}}\cdot \bm{\Sigma}\, \nu^{(-)} = \pm \nu^{(-)}$ for antineutrino, where `$\pm$' stands for the right- and left-helicity states, respectively. This split of energies is discussed in Appendix \ref{sec:A}.  
In the massless limit, i.e., $\widehat{M}^{}_{\nu} = 0$, the difference between the vector and axial vector interactions vanishes for the active neutrino $\nu^{}_{\rm L}$, because the left- and right-handed fields are decoupled.
More technical details on the derivation of the dispersion relation and how the neutrino should be canonically quantized in the dark background are presented in  Appendices \ref{sec:A} and \ref{sec:B}. 
%The structure of this work is organized as follows. In Sec.~2, we first summarize the existing bounds on the considered scenarios from various experiments, and it turns out that the room of the DM effect is large. We then study the effect in $0\nu\beta\beta$ decays in details. In Sec.~3, we display the constraints set by the null signal of the existing $0\nu\beta\beta$-decay experiments, and the sensitivity of future projects is also explored.

\section{\label{sec:3}Model realization}
\noindent
Even though our results derived from KATRIN will be presented in a model-independent form, we want to discuss a specific model realization here for completion.
In general, there are several ways to produce dark potentials for neutrinos in the literature, including the following.
\begin{itemize}[noitemsep,topsep=0pt,leftmargin=5.5mm]
	\item A background of ultralight dark matter, dark radiation or dark energy coupled to neutrinos. The ultralight field can be treated as a classical one, and the neutrino dispersion relation is affected simply by assigning an  expectation value to the dark field in the Lagrangian, e.g., $g^{}_{\phi} \langle {\phi} \rangle \overline{\nu}\nu $ for an ultralight scalar.
	Modified neutrino oscillations in such scenarios have been discussed in detail in the literature~\cite{Berlin:2016woy,Brdar:2017kbt,Krnjaic:2017zlz,Liao:2018byh,Capozzi:2018bps,Reynoso:2016hjr,Huang:2018cwo,Pandey:2018wvh,Farzan:2018pnk,Choi:2019ixb,Baek:2019wdn,Choi:2019zxy,Choi:2020ydp,Dev:2020kgz,Baek:2020ovw,Losada:2021bxx,Smirnov:2021zgn,Alonso-Alvarez:2021pgy}.
	For the scalar ultralight dark matter, the field evolves in the Universe as $\phi = \hat{\phi} \cos{m^{}_{\phi}t}$ with $m^{}_{\phi}$ being the mass of the dark matter field. The field strength reads $\hat{\phi} = \sqrt{2\rho}/m^{}_{\phi}$, where $\rho$ is the energy density of dark matter. For the vector dark matter $A'^{}_{\mu}$, the spatial component $\bm{A}'^{}_{\mu}$ will be averaged out if the polarization of dark matter is randomized, while the temporal component $A'^{}_{0}$ is simply vanishing for free vector particles nearly at rest. 
	\item Coherent forward scattering of neutrinos with massive dark matter particles, the so-called ``dark NSI''~\cite{Ge:2019tdi,Chao:2020qpe}. In this case, it is the elastic scattering of the neutrino wave function off the dark matter grid which changes the dispersion relation. The form of the dark potential depends on the type of  interaction between dark matter $\chi$ and neutrinos.
	However, due to the smallness of the dark matter density, a large coupling and a small mediator mass would be required to achieve an observable potential.
	
	\item A  fifth force sourced by heavy dark matter~\cite{Davoudiasl:2018hjw,Chao:2020qpe} or by ordinary matter~\cite{Joshipura:2003jh,Grifols:2003gy,Grifols:2003gy,Bandyopadhyay:2006uh,Heeck:2010pg,Wise:2018rnb,Ge:2018uhz,Smirnov:2019cae,Babu:2019iml,Bustamante:2018mzu,Coloma:2020gfv,Ge:2021lur}.
	This can be regarded as a special case of coherent forward scattering, as the mediator mass is extremely small, such that the interaction can be described by a long-range force.
	The fifth force as a classical virtual field is able to directly modify the neutrino propagation, similar to the motion of electrons in a Coulomb potential.	Note that to avoid severe constraints from the charged lepton sector, such force is usually assumed to be generated by mixing with sterile neutrinos.

	%For instance, suppose that there is a long-range force $g^{}_{\chi} A'^{}_{\mu} \overline{\chi}\gamma^{\mu}\chi + g^{}_{\nu} A'^{}_{\mu} \overline{\nu}\gamma^{\mu}\nu$ or $g^{}_{\nu} A'^{}_{\mu} \overline{\nu} \gamma^{\mu}\gamma^{}_{5}\nu$  between $\nu$ and the fermionic dark matter $\chi$ mediated by a dark photon $A'$ (similar for a scalar).  The induced dark matter self-interaction may also be helpful to solve the small-scale structure issues. 
	%Similar results can be found for the Yukuwa interaction but with a potential $\langle \phi \rangle \overline{\nu}\nu$~\cite{Davoudiasl:2018hjw}.
\end{itemize}
For illustration, we will elaborate on the last scenario, which can easily realize all forms of dark potentials described in Eq.~(\ref{eq:L}).

In order to generate different forms of the dark potential  in Eq.~(\ref{eq:L}), we consider a fermionic dark matter $\chi$ to source a long-range scalar or vector potential.
The Lagrangian is then given as
\begin{equation}
-\mathcal{L} \supset g^{}_{\chi } A^{\prime}_{\mu} \overline{\chi}\gamma^{\mu}\chi + g^{}_{\chi} \Phi \overline{\chi}\chi \;. 
\label{eq:LDM}
\end{equation}
After integrating out dark matter configurations, we can obtain the fifth force potential sourced by the ambient dark matter. For the vector potential, one ends up with an effective potential $\langle A'^{}_{\mu} \rangle = (g^{}_{\chi} n^{}_{\chi}/m^{2}_{A'},0,0,0)$~\cite{Wise:2018rnb,Smirnov:2019cae} with $n^{}_{\chi}$ being the dark matter number density and $m^{}_{A'}$ being the mediator mass, and the spatial components are vanishing because $\chi$ is non-relativistic.
For the scalar potential, one simply has $\langle \Phi \rangle = g^{}_{\chi } n^{}_{\chi}/m^{2}_{\Phi}$.
With the scalar or vector fifth force, one can obtain the dark potential forms in Eq.~(\ref{eq:L}) by noting $\phi =  \langle \Phi \rangle$,  $\varphi = \langle \Phi \rangle$,  $V^{}_{\mu} =  \langle A^{\prime}_{\mu} \rangle$, $a^{}_{\mu} =  \langle A^{\prime}_{\mu} \rangle $ or $T^{}_{\mu \nu} = \partial^{}_{\mu} \langle A^{\prime}_{\nu} \rangle - \partial^{}_{\nu} \langle A^{\prime}_{\mu} \rangle $.

The Debye screening effect will reduce the magnitude of dark potentials in astrophysical environments with dense mobile $\nu \overline{\nu}$ pairs~\cite{Davoudiasl:2018hjw,Dolgov:1995hc,Esteban:2021ozz}.
In analogy to the electromagnetic screening in a metal with free electrons, the free neutrino pairs will shield the vector field. This is equivalent to giving the vector field a Debye screening mass $m^{}_{\rm DS}$, which causes drastic exponential decrease in the field strength, e.g., $V^{}_{0} \propto \mathrm{e}^{- m^{}_{\rm DS} r}/r$ with $r$ being the distance from the source.
Taking the vector potential for example,
the resultant potential in the solar system with the screening effect reads
\begin{eqnarray}
|g^{}_{\nu} V^{}_{0}| = \frac{g^{}_{\nu} g^{}_{\chi} n^{}_{\chi}}{m^2_{A'} + m^2_{\rm DS}} & \approx & 1.1~{\rm eV} \left(\frac{\rho^{}_{\chi}}{0.3~{\rm GeV\cdot cm^{-3}}} \right) \left( \frac{7~{\rm keV}}{m^{}_{\chi}} \right) \left( \frac{0.01 g^{}_{\chi}}{ g^{}_{\nu}}\right) \frac{(1.95~{\rm K})^2}{ T^{2}_{\nu} + m^2_{A'}/g^2_{\nu}} \; ,
\label{eq:A7}
\end{eqnarray}
where $n^{}_{\chi}$ denotes the local DM density, and the screening mass induced by neutrino plasma is roughly $m^{}_{\rm DS} \sim g^{}_{\nu} T^{}_{\nu}$.
The long-range force parameters can be set to be very small, e.g., $m^{}_{A'} = 10^{-21}~{\rm eV}$ and $g^{}_{\nu} = 10^{-17}$, such that no laboratory bounds other than beta decays can apply to the parameter space of interest. 
One might be concerned about  astrophysical limits from the likes of BBN, CMB, LSS and supernovae, where the dark matter or neutrino density is much higher than in the Earth. However, various arguments attempting to constrain this type of force are mostly invalid due to the screening effect in the dense environment; for relevant details see Refs.~\cite{Davoudiasl:2018hjw,Dolgov:1995hc,Esteban:2021ozz}.

The Lagrangian in Eq.~(\ref{eq:LDM}) will also induce the DM self-interaction.
Before we discuss beta decays, let us briefly comment on the consequences of this self-interaction.
The self-interacting DM is in fact favored by observations of small scale structures~\cite{Bullock:2017xww,Tulin:2017ara}, which itself is a well-motivated topic. The major constraint should come from the observations of structure formation. To avoid making DM too collisional, one puts $g^{}_{\chi} \lesssim 4 \times 10^{-3}\ (m^{}_{\chi}/{\rm GeV})^{3/4}$~\cite{Lasenby:2020rlf}.
There is also a collective bound on  DM long-range interactions from tidal stream of the Sagittarius satellite~\cite{Kesden:2006vz,Kesden:2006zb,Carroll:2008ub}, which has excluded $g^{}_{\chi}/m^{}_{\chi} > 10^{-19}~{\rm GeV}^{-1} $ with $m^{}_{\Phi} \lesssim 4 \times 10^{-28}~{\rm eV}$ corresponding to the Sagittarius dwarf galaxy orbit $16~{\rm kpc}$. 
However, we have checked that most of the parameter space that can give $\mathcal{O}(\rm eV)$ dark potentials remains unconstrained by those considerations.

%Such a bound is reasonable for attractive Yukawa forces and does not apply to the repulsive case in the vector scenario.

\section{The beta-decay rate in the dark sea}
\noindent
The microscopic nature of the nuclear beta decay related to electron neutrinos makes it an excellent complementary probe of the dark potential other than neutrino oscillations.
The amplitude for the transition of beta decays, e.g.\ ${}^3{\rm H} \to {}^3{\rm He} + e^- + \overline{\nu}^{}_{e}$, is given by 
\begin{eqnarray}
% \label{eq:}
\mathcal{M} & = & \frac{G^{}_{\rm F} V^{}_{\rm u d}}{\sqrt{2}} \overline{u}(p^{}_{e}) \gamma^{\mu}_{} (1-\gamma^{}_{5}) v(p^{}_{\nu}) \times \overline{u}(p') \gamma^{}_{\mu} (\bar g^{}_{\rm V}- \bar g^{}_{\rm A}\gamma^{}_{5}) v(p)  \;,
\end{eqnarray}
where $\bar g^{}_{\rm V}$ and $\bar g^{}_{\rm A}$ stand for the vector and axial-vector coupling constants of the charged-current weak interaction of nucleons, respectively, the higher order magnetic and pseudoscalar form factors of the nucleons are neglected, and $p^{}_{}$, $p'$, $p^{}_{e}$ and $p^{}_{\nu}$ are the momenta of  tritium, helium, electron and neutrino, respectively. 
Summing over the spins of particles other than neutrino, we arrive at
\begin{eqnarray}
% \label{eq:}
\sum^{}_{s,s',s^{}_{e}}\left| \mathcal{M} \right|^2 & = & \frac{G^{2}_{\rm F} |V^{}_{\rm u d}|^2}{2} 
{\rm Tr}\left[ (\slashed{p}^{}_{e} + m^{}_{e}) \gamma^\mu (1-\gamma^{}_{5})  v(p^{}_{\nu}, s^{}_{\nu}) \overline{v}(p^{}_{\nu},s^{}_{\nu})  \gamma^\nu (1-\gamma^{}_{5}) 
\vphantom{\slashed{p}^{}_{e}} \right] \notag\\
& & \times  {\rm Tr} \left[\vphantom{\slashed{p}^{}_{e}} (\slashed{p}^{\prime} + M') \gamma^{}_{\mu} (\bar g^{}_{\rm V}- \bar g^{}_{\rm A} \gamma^{}_{5})  (\slashed{p}^{} + M) \gamma^{}_{\nu} (\bar g^{}_{\rm V}- \bar g^{}_{\rm A} \gamma^{}_{5})
\vphantom{\slashed{p}^{}_{e}}\right]  ,
\end{eqnarray}
where for the unsummed spin bilinear of neutrinos $v(p^{}_{\nu}, s^{}_{\nu}) \overline{v}(p^{}_{\nu},s^{}_{\nu})$ under the impact of dark potentials, Eqs.~(\ref{eq:uubarav}) and (\ref{eq:uubarv}) in the Appendix should be taken. 

For the vector dark background, we are ready to sum over the final neutrino spin, i.e.\ $\sum {v}(\bm{p},s)\overline{v}(\bm{p},s) = \slashed{\widetilde{p}}  -  m$.
But for the axial-vector case, due to the split of energy levels discussed in Appendix \ref{sec:A}, the integration over phase space for two helicity states should be performed separately. This introduces extra complexity. 
After the index contraction, the matrix element for the outgoing neutrino with $p^{}_{\nu,s}$ in the axial-vector background is
\begin{eqnarray}
% \label{eq:}
\sum^{}_{s,s',s^{}_{e}} |& \mathcal{M}^{}_{a} & |^2  \approx  16\, G^{2}_{\rm F} |V^{}_{\rm u d}|^2 \left\{ (\bar g^{}_{\rm V} + \bar g^{}_{\rm A})^2 (p^{}_{e} \cdot p')  \vphantom{\slashed{p}^{}_{e}}\right.  \times   (p \cdot p^{}_{\nu,s} ) + (\bar g^{}_{\rm V} - \bar g^{}_{\rm A})^2  (p^{}_{e} \cdot p) (p' \cdot p^{}_{\nu,s} ) \notag\\
& + &  (\bar g^{2}_{\rm A} - \bar g^{2}_{\rm V}) M \cdot M' (p^{}_{e}\cdot p^{}_{\nu,s}) +  2 s^{}_{\nu} \, M M' \left[ (3 \bar g^{2}_{\rm A} + \bar g^{2}_{\rm V}) E^{}_{e} |\bm{p}^{}_{\nu}| +  ( \bar g^{2}_{\rm V} - \bar g^{2}_{\rm A}) E^{}_{\nu,s} \, \bm{p^{}_{e}} \cdot \hat{\bm{p}}^{}_{\nu}  \right]   \notag\\
& - &  \left. g^{}_{a} a^{}_{0} M M'
\left[  (3 \bar g^{2}_{\rm A}+ \bar g^2_{\rm V}) E^{}_{e} 
+  2s^{}_{\nu}(\bar g^{2}_{\rm V} - \bar g^2_{\rm A})  \bm{p^{}_{e}} \cdot \hat{\bm{p}}^{}_{\nu}   \right]  \vphantom{\slashed{p}^{}_{e}} \right\}   .
\end{eqnarray}
With a vanishing $a^{}_{0}$, the result will be reduced to the standard one, which is consistent with Ref.~\cite{Long:2014zva}.
The matrix element in the vector background has a similar expression, but one can sum over the neutrino helicity, leading to
\begin{eqnarray}
% \label{eq:}
\sum^{}_{s,s',s^{}_{e}, s^{}_{\nu}} |& \mathcal{M}^{}_{V} & |^2  =  32 \, G^{2}_{\rm F} |V^{}_{\rm u d}|^2 \left\{ (\bar g^{}_{\rm V} + \bar g^{}_{\rm A})^2 (p^{}_{e} \cdot p')  \vphantom{\slashed{p}^{}_{e}}\right. 
\times (p \cdot \widetilde{p}^{}_{\nu} ) + (\bar g^{}_{\rm V} - \bar g^{}_{\rm A})^2  (p^{}_{e} \cdot p) (p' \cdot \widetilde{p}^{}_{\nu} ) \notag\\
& + &  \left. (\bar g^{2}_{\rm A} - \bar g^{2}_{\rm V}) M \cdot M' (p^{}_{e} \cdot \widetilde{p}^{}_{\nu})  \vphantom{\slashed{p}^{}_{e}}\right\}  ,
\end{eqnarray}
which is close to the standard results but with $p^{}_{\nu}$ in vacuum replaced by $\widetilde{p}^{}_{\nu} = (\widetilde{E}^{}_{\nu}, \bm{p}^{}_{\nu})$.

The final beta-decay rate without the sum of neutrino helicity reads
\begin{eqnarray}
% \label{eq:}
\Gamma^{}_{\beta} & = & \frac{1}{2^9 \pi^5 M} \int \frac{\mathrm{d}^3 \bm{p}' \mathrm{d}^3 \bm{p}^{}_{e} \mathrm{d}^3 \bm{p}^{}_{\nu}}{E' E^{}_{e} \widetilde{E}^{}_{\nu}} \left(\frac{1}{2}\sum^{}_{s,s',s^{}_{e}} \left| \mathcal{M} \right|^2 \right) \times F(Z,E^{}_{e}) \delta^4\left( p - p' - p^{}_{e} - p^{}_{\nu}\right)  .
\end{eqnarray}
Note that the neutrino energy in the phase space factor is different from the vacuum case, namely $\widetilde{E}^{}_{\nu}$ for the vector background and $\widetilde{E} = E^{}_{s}$ for the axial-vector one, such that the normalization and completeness relations of spinors can appreciate the simple forms of Eqs.~(\ref{eq:uuav}), (\ref{eq:uubarav}), (\ref{eq:uuv}) and (\ref{eq:uubarv}) in the Appendix.
In principle, one can take different normalization conventions, but the final result is invariant. The neutrino energy in the delta function should take the form in Eq.~(\ref{eq:Es}) or (\ref{eq:Epm}).

The integration should be done in the rest frame of the tritium, in accordance with the frame picked out by the dark sea considering the Earth is non-relativistic.
After the trivial integration over $\mathrm{d}^3 \bm{p}$ and decomposing $\mathrm{d}^3 \bm{p}^{}_{e} = |\bm{p}^{}_{e}|^2 \mathrm{d} \bm{p}^{}_{e} \mathrm{d} \cos{\theta^{}_{e\nu}} \mathrm{d} \phi^{}_{e\nu}$, we have
\begin{eqnarray}
% \label{eq:}
\Gamma^{}_{\beta} & = & \frac{1}{2^8 \pi^4 M} \int 
\frac{  |\bm{p}^{}_{e}|^2 \mathrm{d} |\bm{p}^{}_{e}| \mathrm{d} \cos{\theta^{}_{e\nu}} \cdot
\mathrm{d}^3 \bm{p}^{}_{\nu}}{E' E^{}_{e} \widetilde{E}^{}_{\nu}}     \left(\frac{1}{2}\sum^{}_{s,s',s^{}_{e}} \left| \mathcal{M} \right|^2 \right) F(Z,E^{}_{e})  \delta\left( E - E' - E^{}_{e} - E^{}_{\nu}\right)  .
\end{eqnarray}
Since
$\mathrm{d} \cos{\theta^{}_{e\nu}} = E' / (|\bm{p}^{}_{e}|\cdot |\bm{p}^{}_{\nu}|) \mathrm{d} E'$ and the neutrino favors no specific direction,
the decay rate is simplified to 
\begin{eqnarray}
\label{eq:Gb}
\Gamma^{}_{\beta} & = & \frac{1}{2^6 \pi^3 M} \int 
\frac{  \mathrm{d} E^{}_{e}\cdot |\bm{p}^{}_{\nu}|
\mathrm{d}|\bm{p}^{}_{\nu}|}{ \widetilde{E}^{}_{\nu}  }   \left(\frac{1}{2}\sum^{}_{s,s',s^{}_{e}} \left| \mathcal{M} \right|^2 \right) F(Z,E^{}_{e})  .
\end{eqnarray}
We are left with integrating over the neutrino momentum in order to obtain the differential spectrum with respect to the electron energy.
The integration limit of $|\bm{p}^{}_{\nu}|$ can be obtained by requiring that 
\begin{eqnarray}
% \label{eq:}
E^{}_{\nu}(|\bm{p}^{}_{\nu}|) + E^{}_{e}(|\bm{p}^{}_{e}|) + E'(|\bm{p}^{}_{\nu}|,|\bm{p}^{}_{e}|, \cos{\theta^{}_{e\nu}}) = E
\end{eqnarray}
has a solution for any $-1 \leqslant \cos{\theta^{}_{e\nu}} \leqslant 1$ and $ m^{}_{e} \leqslant E^{}_{e} \leqslant E^{\rm  max}_{e}$. Trivial analytical solutions exist for the standard~\cite{Masood:2007rc} and vector cases.
For the axial-vector case, we integrate the rate numerically.
It is worthwhile to remark that at the maximal electron energy (i.e., minimal neutrino energy) in the axial-vector case, $|\bm{p}| = 2s g^{}_{a} a^{}_{0}$, and the phase space factor in Eq.~(\ref{eq:Gb}) is not vanishing as in the standard case.  This gives rise to a finite decay rate at the endpoint of electron spectrum.
%A simplified spectrum can be obtained by considering the accuracy pursued by the KATRIN experiment. These are

\section{Signals at KATRIN}

%One of the straightforward effects of Eq.~(\ref{eq:dispersion}) is the distortion of the beta-decay spectrum.
%Before going into details, let us first remark on the effects of different potentials based on simple observations from Eq.~(\ref{eq:dispersion}). 
%The vector interaction will shift the whole beta-decay spectrum, including the endpoint. The axial-vector interaction is similar to a mass term at vanishing momentum, but the dispersion relation scales in a completely different way when the momentum is non-vanishing, which should induce a clear distortion to the original beta spectrum.

In the absence of dark field, the rate of beta decays, ${}^3{\rm H} \to {}^3{\rm He} + e^- + \overline{\nu}^{}_{e}$, reads~\cite{Shrock:1980vy,Masood:2007rc,Simkovic:2007yi,Long:2014zva,Ludl:2016ane}
\begin{eqnarray}\label{eq:betaExact}
\frac{\mathrm{d} \Gamma^{}_{\rm \beta}}{\mathrm{d} K^{}_{e}} & = & N^{}_{\rm T} \frac{\overline{\sigma}(E^{}_e)}{\pi^2} \sum^3_{i=1} |U^{}_{ei}|^2 H(E^{}_{e},m^{}_{i})     ,
%     (3)
\end{eqnarray}
where $N^{}_{\rm T}$ is the total mass of the tritium sample, and $\overline{\sigma}(E^{}_e)$ is the reduced cross section (see, e.g., Ref.~\cite{Huang:2019tdh} for the expression).  
%where $G^{}_{\rm F} = 1.166\times 10^{-5}~{\rm GeV}^{-2}$, $|V^{}_{\rm ud}| \approx \cos \theta^{}_{\rm C}$ with $\theta^{}_{\rm C} \approx 12.8^\circ$,  
%$F(Z, E^{}_e) = 2\pi \eta/(1 - e^{-2\pi \eta})$ is the ordinary Fermi function, $\eta \equiv Z \alpha E^{}_e/p^{}_e$ with $\alpha \approx 1/137$, and $\bar g^{}_{\rm V} \approx 1$ and $\bar g^{}_{\rm A} \approx 1.247$ stand for the vector and axial-vector coupling constants of weak interactions of tritium, respectively.
The kinematics of the beta-decay spectrum is contained in (defining $K_e = E_e - m_e$) 
\begin{eqnarray}\label{eq:Hfunction}
H (E^{}_{e}, m^{}_{i}) & \approx &  \sqrt{(K^{}_{\rm end,0}-K^{}_{e})^2 - m^{2}_{i}} \times  \left(K^{}_{\rm end,0} - K^{}_{e}\right)    ,\notag
\end{eqnarray}
where $K^{}_{\rm end,0}= \left[(m^{}_{^3{\rm H}}-m^{}_{e})^2-m^{2}_{^3{\rm He}}\right]/(2m^{}_{^3{\rm H}})$ is the endpoint energy in the relativistic theory assuming a vanishing neutrino mass. 
The actual endpoint energy for the neutrino mass $m^{}_{i}$
is approximately given by 
\begin{eqnarray}\label{eq:endvac}
K^{}_{\rm end} = K^{}_{\rm end,0} - m^{}_{i}\;.
\end{eqnarray}
%where the last factor $ {m^{}_{^3 {\rm  He}}}/{m^{}_{^3 {\rm H}}} \approx 1$ is important for the far future beta-decay experiment PTOLEMY~\cite{Huang:2019tdh}. In this work, we focus on the KATRIN experiment, for which the endpoint energy is safely taken to be $K^{}_{\rm end} = K^{}_{\rm end,0} - m^{}_{i}$.

As mentioned above, the scalar and pseudoscalar potentials merely add an effective mass term to neutrinos, which is kinetically indistinguishable from the vacuum mass. In fact, in some scenarios they are even postulated to be the origin of small neutrino masses~\cite{Davoudiasl:2018hjw}. 
However, we need to emphasize that
since the current dark potential is expected to be different from that in the early Universe, the model-dependent cosmological bounds on the absolute neutrino mass, e.g.\  $\Sigma < 0.12~{\rm eV}$~\cite{Planck:2018vyg}, can be evaded or weakened. This will possibly lead to  large signals in future KATRIN runs, which expect no visible effect of neutrino masses if the stringent cosmological bounds are adopted~\cite{Huang:2019tdh}.
%In other words, if future KATRIN runs observe a finite neutrino mass, it might be attributed to (pseudo)scalar dark potentials.
%Second, if the interaction with dark matter is the source of such potentials, the neutrinos in the cosmic voids will process different mass and mixings. This may alter, for instance, the flavor ratio of ultrahigh energy neutrinos~\cite{}.
%Third, the ultralight dark matter scenario can induce modulation signals rather than a constant spectrum shape; see Refs.~\cite{} for discussions on modulation effect of neutrino oscillation experiments. Last, 

The vector potential has a profound effect on the beta-decay spectrum. To clearly see that, we pick out from Eq.~(\ref{eq:dispersion}) the vector contribution to the energy, which reads
$
E^{}_{\nu} =  \sqrt{\bm{p}^{2}_{\nu} + m^2_{i}} + g^{}_{V} V^{}_{0} 
$
for the neutrino mode and 
$
E^{}_{\nu} =  \sqrt{\bm{p}^{2}_{\nu} + m^2_{i}} - g^{}_{V} V^{}_{0} 
$
for the antineutrino mode. 
The fact that  neutrino and antineutrino excitations feel opposite vector potentials implies that neutrino mass experiments using electron capture such as ECHo \cite{Gastaldo:2017edk}, will see an opposite effect compared to KATRIN, providing a way to independently test the effect. 
The antineutrino energy of beta decays in the vector background can run into the negative (but bounded), when $g^{}_{V} V^{}_{0} > \sqrt{\bm{p}^{2}_{\nu} + m^2_{i}}$. The beta-decay spectrum can hence extend beyond the normal kinematic limit $K^{}_{\rm end,0}$.
This is not a surprise as
the process which is not kinematically allowed in vacuum can take place if the medium modifies the dispersion relations~\cite{Raffelt:1996wa}, a phenomenon familiar in, e.g., plasmon decay.
%This is also very similar to the pair production of electron and positron in an electrostatic field.
%The beta decay products, i.e., electron and helium, can gain energy larger than the tritium mass, at the cost of creating a neutrino bounded with the vector background. 
%Another analogy is that as we calculate the beta spectrum, the effect of nuclear coulomb potential will distort a bit the electron wavefunction via $F(Z, E^{}_e)$.
%From this point of view, the term $g^{}_{V} V^{}_{0}$ should be interpreted as the binding energy of the neutrino-vector system.
In the presence of the vector potential, the electron endpoint energy in Eq.~(\ref{eq:endvac}) will be shifted towards 
\begin{eqnarray}% \label{eq:}
\widetilde{K}^{}_{\rm end}
&  \approx & K^{}_{\rm end,0} - m^{}_{i} + g^{}_{V} V^{}_{0} \; .
\end{eqnarray}
Since the axial-vector interaction distinguishes two helicity states, to exactly calculate the spectrum we have to perform the integration over the unsummed helicity amplitudes with Eq.~(\ref{eq:Gb}).

%%%%%%%%%%%%%%%%%%%%%%%%%%%%%%%%%%%%%%%%%%%%%%%%%%%
\begin{figure}[t!]
	\begin{center}
		\vspace{-0.3cm}
		\includegraphics[width=0.48\textwidth]{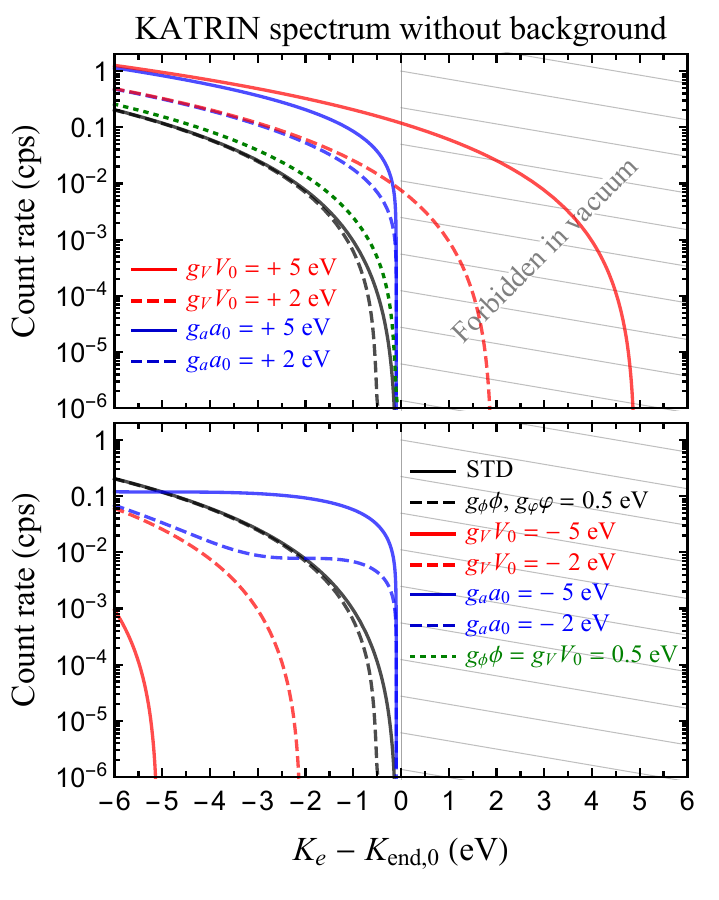}
		\includegraphics[width=0.48\textwidth]{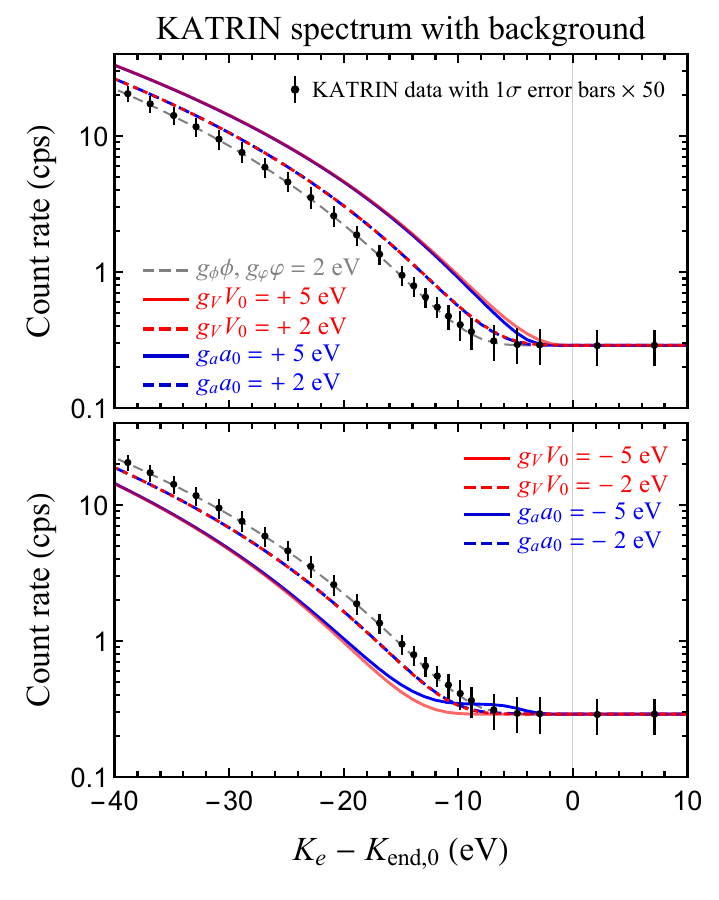}
	\end{center}
	\vspace{-0.3cm}
	\caption{The integrated beta-decay spectra for the first KATRIN campaign in various types of dark seas:  vector potentials with $g^{}_{V}V^{}_{0} = \pm 5~{\rm eV}$ (solid red curves) and $g^{}_{V}V^{}_{0} = \pm 2~{\rm eV}$ (dashed red curves) as well as  axial-vector potentials with $g^{}_{a}a^{}_{0} = \pm 5~{\rm eV}$ (solid blue curves) and $g^{}_{a}a^{}_{0} = \pm 2~{\rm eV}$ (dashed blue curves). 
		%The standard case without any dark potential effects is shown as the solid black curve. 
		For these curves, the neutrino mass has been fixed to $m^{}_{1}=0.1~{\rm eV}$. Contributions from $\phi$ and $\varphi$ which mimic a neutrino mass exceeding cosmological bounds are shown as the dashed gray curve. The left (right) panel gives the spectra without (with) the background. In the right panel, the KATRIN data points are also given for comparison. Note that for demonstration, the errors are shown as fifty times the standard deviation.}
	\label{fig:beta}
\end{figure}
%%%%%%%%%%%%%%%%%%%%%%%%%%%%%%%%%%%%%%%%%%%%%%%%%%%	

%In the relativistic limit (i.e., away from the endpoint), only the outgoing right-helicity antineutrino is active, and the vector and axial-vector cases should converge to each other. 
%Note that  for the axial-vector case electron capture experiments will also see an opposite effect compared to KATRIN. 

We assume ${\rm T}^{}_{2}$ as the only tritium source, and only the  final-state excitations of ${}^3{\rm He T}^+$ need to be considered. 
The differential spectrum will have to sum over the final-state distributions of the daughter molecule.
Given the accuracy of KATRIN, we use the Gaussian-averaged final-state distributions in Ref.~\cite{Doss:2006zv}.
The ultimate integrated event rate at KATRIN is given by convolution of the differential spectrum ${\mathrm{d} \Gamma^{}_{\rm \beta}}/{\mathrm{d} K^{}_{e}}$ with the spectrometer response function. 
The predicted rate is given by 
\begin{eqnarray}
% \label{eq:}
R^{}_{\rm th} = A^{}_{\rm s} N^{}_{\rm T} \int f^{}_{\rm res}( K^{}_{e} - qU ) \frac{\mathrm{d}\Gamma^{}_{\beta}}{\mathrm{d} K^{}_{e}} \mathrm{d} K^{}_{e} + R^{}_{\rm bkg}  \;,
\end{eqnarray}
where $A^{}_{\rm s}$ is the normalization factor, $N^{}_{\rm T}$ is the target tritium number, $qU$ is the applied retarding potential, and $R^{}_{\rm bkg}  $ is the background rate.  
Here, we have assumed a constant background rate within the energy window of interest, following the KATRIN collaboration~\cite{Aker:2021gma}. If the vector potential is too large, the reconstruction of the background rate can be affected. However, we note that the background and signal follow different spectrum shapes and can be statistically distinguished by the fitting procedure.
%The variables $A^{}_{\rm s}$, $R^{}_{\rm bkg}$ and $m^2_{\nu}$ are taken as free parameters. 
The response function $f^{}_{\rm res}( K^{}_{e} - qU )$ is dependent on the surplus energy $E^{}_{e}-q U$, where $qU$ is the applied electric potential.
For the first campaign of KATRIN, we use the response function  given in Fig.~2 of Ref.~\cite{KATRIN:2019yun} with a column density $1.11 \times 10^{17}~{\rm molecules \cdot cm^{-2}}$.
The predicted rate is to be compared with the measured one, for which
the ring-averaged event rate with statistical and systematic errors is available in Ref.~\cite{Aker:2021gma}.
The variables $A^{}_{\rm s}$, $R^{}_{\rm bkg}$ and $m^2_{\nu}$ will be taken as free parameters during the fit.

In Fig.~\ref{fig:beta}, we have illustrated the distortions of beta-decay spectrum in various dark seas at KATRIN. 
The left panel stands for the ideal case without taking into account the background, while the right one gives a more realistic result with the background.
The vector potential (red curves) shifts the whole spectrum to lower or higher endpoints, without changing the spectral shape.
%The kinematically forbidden region of the spectrum is reachable in the case of a dark vector potential, as we discussed previously.
In comparison, the axial-vector potential (blue curves) induces a non-trivial distortion to the spectrum near the endpoint, but it indeed converges to the vector case away from the endpoint.
By measuring this distortion, one can distinguish between the effects of vector and axial-vector potentials. However, we find numerically that current KATRIN runs are not yet sensitive to this distortion, which would require more statistics. We expect that the Project 8 experiment and PTOLEMY proposal can provide better sensitivities to probe such a distortion near the endpoint, but how good the sensitivity is would require a further study.
On the other hand, the scalar potentials $\phi$ and $\varphi$ mimic the effect of  neutrino masses, shown as dashed gray curves. They can induce an effective neutrino mass beyond the cosmological constraint.
 Even though we turn on only one dark potential at a time, in the left panel of Fig.~\ref{fig:beta} we show the case of $g^{}_{\phi}\phi = g^{}_{V}V^{}_{0} = 0.5~{\rm eV}$ as the dotted green curve to imply the existence of degeneracy when multiple dark potentials are turned on. One can notice that the effects of two dark potentials can counterbalance each other to some extent. 

It is worthwhile to mention that the previous anomalous signal detected by Troitsk~\cite{Belesev1995ResultsOT} may be explained by the vector potential, which can shift the endpoint even beyond its maximum value and affect the reconstruction of $m^2_{\nu}$. However, the anomalous signal has disappeared in later measurements~\cite{Aseev:2011dq}. To avoid false signals, experimental systematics must be well controlled in order to probe such new physics effects.

In the limit of  massless neutrinos, the difference between vector and axial-vector dark potentials should be vanishing, as expected from the analysis in Sec.~\ref{sec:2}. However, as in Fig.~\ref{fig:beta} the vector and axial-vector cases have an apparent difference and cannot smoothly converge by taking the neutrino mass to be vanishing.
Only when the neutrino mass is comparable to time-scale of the background field formation (typically $\sim 1~{\rm Gyr}$ corresponding to $10^{-32}$ eV), the axial-vector scenario starts to approach the vector case. For further discussions, see Appendix~\ref{sec:C}.

%%%%%%%%%%%%%%%%%%%%%%%%%%%%%%%%%%%%%%%%%%%%%%%%%%%
\begin{figure}[t!]
\begin{center}
	\includegraphics[width=0.48\textwidth]{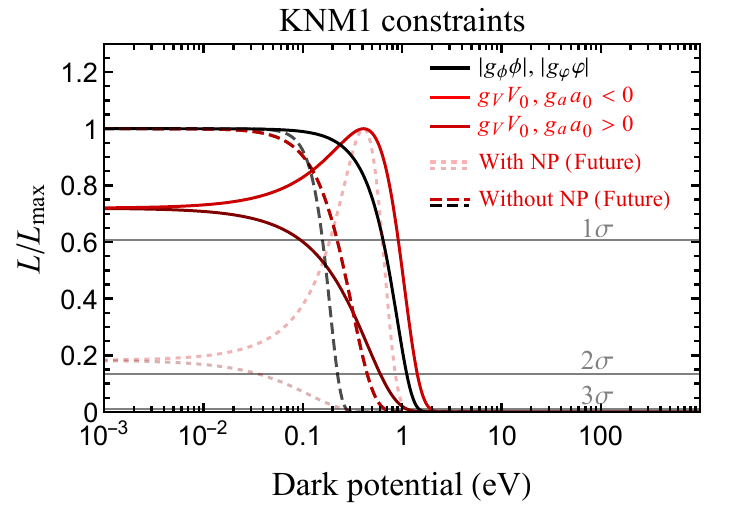}
\end{center}
\vspace{-0.3cm}
\caption{The constraints on the dark potentials $g^{}_{V}V^{}_{0}$, $g^{}_{a}a^{}_{0}$, $g^{}_{\phi} \phi$ and $g^{}_{\varphi}\varphi$ using the data of KATRIN's first campaign, shown as the solid curves. 
We have normalized the likelihood $L$, i.e., dividing by the likelihood maximum $L^{}_{\rm max}$ in each case.	
The sensitivities corresponding to the ultimate KATRIN goal $m^{}_{\nu}<0.2~{\rm eV}$ at $90\%$ level, i.e., $\sigma(m^2_{\nu}) = 0.025~{\rm eV}^2$~\cite{KATRIN:2005fny}, along with a reduced potential fluctuation by a factor of three are given as the dotted lighter curves (in the presence of a potential) and dashed curves (without new physics). 
The horizontal gray lines represent $\Delta \chi^2=1$, $4$ and $9$, respectively. 
	%The maximally allowed values of $g^{}_{0} V^{}_{0}$ and $g^{}_{a} a^{}_{0}$ in the framework of a fifth force sourced by dark matter are shown as the vertical line.
}
\label{fig:constraints}
\end{figure}
%%%%%%%%%%%%%%%%%%%%%%%%%%%%%%%%%%%%%%%%%%%%%%%%%%%

% (it also happens in the tachyon context~\cite{Ciborowski:1998kc})
We continue with fitting KATRIN data to our scenarios. 
The effect of scalar and pseudoscalar potentials is identical to being from  $m^{2}_{\nu}$. 
The consequences of vector and axial-vector potentials are the same at  energies away from the endpoint, for which KATRIN collects the most events.
For the present KATRIN sensitivity, the major effects of the vector and axial-vector potentials are to shift the endpoint energy $E^{}_{0}$ of electrons. Hence their effects are entirely ascribed to the fit of $E^{}_{0} \equiv K^{}_{\rm end}  + g^{}_{V} V^{}_{0}$ at KATRIN.
The endpoint $E^{}_{0}$ and the squared neutrino mass $m^2_{\nu}$ are regarded as free parameters in KATRIN fits~\cite{KATRIN:2019yun,KATRIN:2021fgc,Aker:2021gma}. 
We perform our own fit with available data of the first KATRIN neutrino mass campaign (KNM1)~\cite{KATRIN:2019yun}. 
For this purpose, we vary freely the vacuum neutrino mass over $m^{2}_{\nu}\geq 0$.
In the official fit of KATRIN, the $m^2_{\nu}<0$ region is kept to account for data fluctuations, but eventually it is removed by proper statistical  interpretations in obtaining the mass limit. We adopt here a simplified approach with the main interest being the minimized $\chi^2$, whose function for KNM1 is constructed as
\begin{equation}
\chi^2_{\beta} (m^2_{\nu},E^{}_{0}, A^{}_{\rm s}, R^{}_{\rm bkg}) = \frac{(R^{i}_{\rm th} - R^{i}_{\rm exp})^2}{(\sigma^{i}_{\rm sta})^2 + (\sigma^{i}_{\rm sys})^2} \; , 
\end{equation}
where the statistical and systematic uncertainties, $\sigma^{i}_{\rm sta}$ and $\sigma^{i}_{\rm sys}$, are taken from  Ref.~\cite{Aker:2021gma} for each retarding potential $q U^{}_{i}$. For KNM1, there are 27 set points of $q U^{}_{i}$ in total, which have already been shown in the right panel of Fig.~\ref{fig:beta}.
We hence assume the vacuum neutrino mass to be vanishing  and vary other parameters $E^{}_{0}$, $A^{}_{\rm s}$ and $R^{}_{\rm bkg}$ freely. 
Marginalizing over parameters other than $E^{}_{0}$~\footnote{In order to obtain the constraint on the endpoint energy of interest, we should minimize $\chi^2$ by scanning over all the other parameters  including $A^{}_{\rm s}$, $R^{}_{\rm bkg}$ and $m^{2}_{\nu}$.}, our fit result on the endpoint will be compared with the expected value, and this is then used to set a limit on $g^{}_{V} V^{}_{0}$. 
Note that we do not attempt to fit the  results of the second KATRIN neutrino mass campaign (KNM2) here~\cite{Aker:2021gma}, for which the statistical framework should account for every detector ring and there is not sufficient information to perform the fit ourselves. 
%
%In order to indicate the constraining power of KNM2, we take the tabulated distribution results of $E^{}_{0}$ and $m^2_{\nu}$.

The fit yields $E^{}_{0} = 18573.79 \pm 0.02~{\rm eV} $. The actual $Q$-value is obtained by correcting for  molecular recoil (1.72~eV) and  potential fluctuations of the tritium source and main spectrometer ($-0.2 \pm 0.5~{\rm eV}$ for KNM1), which gives ${Q} = 18575.31 \pm 0.5~{\rm eV}$, slightly smaller than the  expectation of  $18575.72 \pm 0.07 ~{\rm eV}$. 
For the (pseudo)scalar potential, we assume the vacuum mass to be vanishing.

%For KNM2 data~\cite{Aker:2021gma},  we directly use the official results $E^{}_{0} = 18573.69 \pm 0.03~{\rm eV}$ and $m^{2}_{\nu} = 0.26 \pm 0.34~{\rm eV}^2$ (which is mostly in the positive regime) to generate the $\chi^2$ for our dark potentials.

The likelihood $L = \exp{(-\Delta\chi^2/2)}$ for our dark potentials is given in Fig.~\ref{fig:constraints}.
The $2\sigma$ limits corresponding to $\Delta \chi^2 < 4$ read $ -1.4~{\rm eV} < (g^{}_{V} V^{}_{0},\, g^{}_{a}a^{}_{0}) < 0.6~{\rm eV}$ and $(|g^{}_{\phi} \phi|,\, |g^{}_{\varphi} \varphi|) < 1.1~{\rm eV}$.
For potential values smaller than $\mathcal{O}(1)~{\rm eV}$, the $\chi^2$ is dominated by events away from the endpoint, where the effects of vector and axial-vector become almost the same.
%Note that because in  KATRIN fits the endpoint energy $E^{}_{0}$ is taken to be completely free without any priors, the vector and axial-vector potentials will not affect the current neutrino mass results.

A slight preference of the (axial-)vector potential $g^{}_{V} V^{}_{0} = -0.42~{\rm eV}$ can be noted.
%The current sensitivity of KATRIN is not sufficient to distinguish vector and axial-vector cases. 
Keeping the best-fit values so far, future sensitivities to reach the reference KATRIN target $m^{}_{\nu} < 0.2~{\rm eV}$ and by reducing potential fluctuations by a factor of three are shown as dotted curves. 
In comparison, assuming no new physics contributions, the sensitivities are instead given by the dashed curves.

\section{Conclusion}
We have performed a novel and systematic study on the effects of dark neutrino potentials on  beta decays, especially focusing on the ongoing KATRIN experiment. By collectively interacting with background fields, neutrinos will have dispersion relations different from the ones in vacuum, which induces distinct distortions to the beta-decay spectrum. 
Observable consequences include neutrino mass signals beyond the ones bounded from cosmological constraints, events beyond the kinematical endpoint of the decay, and spectral distortions. 
We find that the current KATRIN data favor the (axial-)vector potential $g^{}_{V} V^{}_{0} = -0.42~{\rm eV}$, but more statistics should be required to draw a more robust conclusion. The current KATRIN runs are not yet sensitive to the discrimination between vector and axial-vector cases, but  future experiments may be able to achieve this by more accurately measuring the distortions near the endpoint.

The ECHo experiment with the electron capture technique will feel an opposite vector potential compared to KATRIN, which can provide a complementary probe if the dark potential is present.
The next-generation beta-decay experiment like Project 8 (with molecular or atomic tritium) measuring the differential spectrum can be an excellent further probe of dark potentials, providing a well controlled energy scale of the spectrometer.
With a tritium source as large as $100~{\rm g}$, we expect the PTOLEMY proposal to have an overwhelmingly better sensitivity, which is interesting for a future work.

\section*{Acknowledgments}
The authors would like to thank Shu-Yuan Guo, Leonard K\"ollenberger, Newton Nath, Kathrin Valerius and Shun Zhou for valuable communications. This work is supported in part by the Alexander von Humboldt Foundation.
%\end{acknowledgements}

%%%%%%%%%%%%%%%%%%%%%%%%%%%%%%%%%%%%%%%%%%%%%%%%%%
\appendix

\appendix

\section{\label{sec:A}Massive neutrinos in an  axial-vector background}
\noindent The plane-wave solutions to the Dirac equations in the axial-vector background differ from the vacuum ones. To see that, we collect the left- and right-handed fields into $\nu = \nu^{}_{\rm L} + \nu^{}_{\rm R}$ ($\nu^{}_{\rm R} = \nu^{\rm c}_{\rm L}$ for Majorana neutrinos), which satisfies the equation of motion
\begin{eqnarray}
(i \slashed{\partial} + g^{}_{a} a^{}_{0} \gamma^{0} \gamma^{5}_{} - m) \nu = 0 \;.
\end{eqnarray}
For the positive and negative frequency modes, we have
\begin{eqnarray}
\label{eq:spinorAxial}
(p^{}_{\mu} \gamma^\mu + g^{}_{a} a^{}_{0} \gamma^{0} \gamma^{5}_{} - m) u(p,s) & =  & 0 \;, \\
(-p^{}_{\mu} \gamma^\mu + g^{}_{a} a^{}_{0} \gamma^{0} \gamma^{5}_{} - m) v(p,s) & = & 0 \;.
\end{eqnarray}
First of all, the energy eigenvalues should be derived. This can be done by multiplying a matrix $(p^{}_{\mu} \gamma^\mu + g^{}_{a} a^{}_{0} \gamma^{0} \gamma^{5}_{} + m)$ from the left to Eq.~(\ref{eq:spinorAxial}), yielding
\begin{eqnarray}
%% \label{eq:}
\left[p^2- (g^{}_{a} a^{}_{0})^2 + 2 g^{}_{a} a^{}_{0}\, \bm{p}\cdot \bm{\Sigma}  - m^2 \right]u(p,s) = 0 \; , \notag
\end{eqnarray}
where $\bm{\Sigma} =\gamma^5 \gamma^0 \bm{\gamma}$ is simply the spin operator, and $\hat{\bm{p}}\cdot \bm{\Sigma}/2$ represents the helicity with eigenvalue $s = \pm 1/2$.
Different from the vacuum case, the neutrino energy is split for the two helicities by the temporal component of the background.
The energy eigenvalues for both neutrino and antineutrino (the same for Majorana case) read
\begin{eqnarray}
\label{eq:Es}
E^{}_{s} = \sqrt{(|\bm{p}| - 2 s\cdot g^{}_{a}a^{}_{0} )^2 + m^2} \; .
\end{eqnarray}
Note that the above equation does not apply to the massless case with $m=0~{\rm eV}$.
By imposing the orthogonality conditions\footnote{One can check that these orthogonality conditions guarantee the plane wave to be normalized under the field expansion Eq.~(\ref{eq:nuxav}), e.g., $\int \mathrm{d}^3 x\,  \nu^\dagger_{p',s'} \nu^{}_{p,s} = \delta^3(p'-p)\delta^{}_{s's}$.}, 
\begin{eqnarray}
& &{u}^\dagger(\bm{p},s') u(\bm{p},s) = {v}^\dagger(\bm{p},s') v(\bm{p},s)= 2 E^{}_{s} \delta^{}_{s's} \;, \notag\\
\label{eq:uuav}
& & {u}^\dagger(-\bm{p},s') v(\bm{p},s) = {v}^\dagger(-\bm{p},s') u(\bm{p},s)= 0 \;,
\end{eqnarray}
the structure of the spinors $u$ and $v$ is found to be of the form
\begin{eqnarray}
% \label{eq:}
u(p,s) & = & \frac{\slashed{p}^{}_{s} + g^{}_{a}a^{}_{0}\gamma^0 \gamma^5+ m  }{\sqrt{(E^{}_{s} + E^{}_{0})^2 - 4(s\, p - g^{}_{a} a^{}_{0})^2}} u(0,s) \;, \notag\\
v(p,s) & = & \frac{-\slashed{p}^{}_{s} + g^{}_{a}a^{}_{0}\gamma^0 \gamma^5+ m  }{\sqrt{(E^{}_{s} + E^{}_{0})^2 - 4(s\, p - g^{}_{a} a^{}_{0})^2}} v(0,s)\; ,  \notag
\end{eqnarray}
where $E^{}_{0} = \sqrt{(g^{}_{a} a^{}_{0})^2 + m^2}$ is the neutrino energy at rest. Notice that for the spinor $v$, we have the relation $\hat{\bm{p}}\cdot \bm{\Sigma}\, v(p,s) = -2 s\, v(p,s)$, in comparison to $\hat{\bm{p}}\cdot \bm{\Sigma}\, u(p,s) = 2 s\, u(p,s)$.
The helicity completeness relations of the spinors are
\begin{eqnarray}
{u}(\bm{p},s)\overline{u}(\bm{p},s) & = &  (\slashed{p}^{}_{s} + g^{}_{a}a^{}_{0}\gamma^0 \gamma^5 +  m )  \frac{1+ 2 s\, \hat{\bm{p}}\cdot \bm{\Sigma}}{2} \; , \notag\\
\label{eq:uubarav}
{v}(\bm{p},s)\overline{v}(\bm{p},s) & = &  (\slashed{p}^{}_{s} - g^{}_{a}a^{}_{0}\gamma^0 \gamma^5  -  m )  \frac{1 - 2 s\, \hat{\bm{p}}\cdot \bm{\Sigma}}{2} \; .
\end{eqnarray}
It is easy to verify these relations with the help of the orthogonality conditions, and one recovers the standard results in the limit of $a^{}_{0}=0$.

Expanding the field operator as $\nu = \int \mathrm{d}^3\bm{p} (b\, \nu^{(+)}_{p,s} + d^\dagger \,\nu^{(-)}_{p,s})$, we arrive at 
\begin{eqnarray}
\label{eq:nuxav}
\nu(x) =   \sum^{}_{s} \int \frac{\mathrm{d}^3 \bm{p}}{(2\pi)^{3/2}} & &\sqrt{\frac{1}{2E^{}_{s}}} \left[  b^{}_{p,s} u(\bm{p},s) {e}^{- i p^{}_{s}\cdot x} + d^\dagger_{p,s} v(\bm{p},s) {e}^{ i p^{}_{s}\cdot x} \right]  ,
\end{eqnarray}
where $b^{}_{p,s}$ and $d^{\dagger}_{p,s}$ should be interpreted as the particle annihilation and antiparticle creation operators, respectively, for Dirac neutrinos. For Majorana neutrinos, the condition $\nu = \nu^{\rm c}_{}$ will force $b^{}_{p,s} = d^{}_{p,s}$, i.e., $b^{}_{p,s}$ annihilates simultaneously the positive- and negative-frequency excitations.
Using the orthogonality conditions, the canonical quantization rules of $\nu(x)$ consistently lead to 
\begin{eqnarray}
% \label{eq:}
\left\{b^{}_{p',s'},b^\dagger_{p,s} \right\} & = & \delta^3(\bm{p}'-\bm{p})\delta^{}_{s's} \; ,\notag \\
\left\{d^{}_{p',s'},d^\dagger_{p,s} \right\} & = & \delta^3(\bm{p}'-\bm{p})\delta^{}_{s's}  \;.
\end{eqnarray}
The neutrino Hamiltonian with normal ordering can then be expanded as (e.g., for the Dirac case)
\begin{eqnarray}
% \label{eq:}
\mathcal{H} =  \int \mathrm{d}^3 \bm{p} \sum^{}_{s} E^{}_{s} \left( b^\dagger_{p,s} b^{}_{p,s} + d^\dagger_{p,s} d^{}_{p,s} \right)   .
\end{eqnarray}

\section{\label{sec:B}Dirac neutrinos in a vector background}
\noindent The results for the vector background are more straightforward. Given the EOM 
\begin{eqnarray}
(i \slashed{\partial} - g^{}_{V} \slashed{V} - m) \nu = 0 \;,
\end{eqnarray}
the plane-wave spinor should satisfy
\begin{eqnarray}
% \label{eq:}
\left[ (p^{}_{\mu} - g^{}_{V}  V^{}_{\mu}) \gamma^\mu - m) \right] u(p,s) & =  & 0 \;, \\
\left[ (-p^{}_{\mu} -  g^{}_{V}  V^{}_{\mu}) \gamma^\mu - m) \right] v(p,s) & = & 0 \;.
\end{eqnarray}
This leads to the energy eigenvalues
\begin{eqnarray}
\label{eq:Epm}
E^{}_{\pm} = \sqrt{(\bm{p} \mp g^{}_{V}\bm{V})^2 + m^2} \pm g^{}_{V}V^{}_{0} \; ,
\end{eqnarray}
where ``$\pm$'' in $E^{}_{\pm}$ corresponds to the neutrino ($u$) and the antineutrino ($v$), respectively. Note again that these results do not apply to Majorana neutrinos. Hence, in principle one can distinguish Majorana and Dirac neutrinos by the experimental signature of the vector potential, if one would know that the interaction is diagonal in flavor. 

The orthogonality conditions as well as the completeness relations are consistently given by 
\begin{eqnarray}
\label{eq:uuv}
{u}^\dagger(\bm{p},s') u(\bm{p},s) & = & {v}^\dagger(\bm{p},s') v(\bm{p},s)=2\widetilde{E} \delta^{}_{s's}\; , \notag\\
{u}^\dagger(-\bm{p},s') v(\bm{p},s) & = & {v}^\dagger(-\bm{p},s') u(\bm{p},s)= 0 \;,\\
{u}(\bm{p},s)\overline{u}(\bm{p},s) & = &  (\slashed{\widetilde{p}}  +  m )  \frac{1+ 2 s\, \hat{\bm{p}}\cdot \bm{\Sigma}}{2}  \;, \notag\\
\label{eq:uubarv}
{v}(\bm{p},s)\overline{v}(\bm{p},s) & = &  (\slashed{\widetilde{p}}  -  m ) \frac{1 - 2 s\, \hat{\bm{p}}\cdot \bm{\Sigma}}{2} \; ,
\end{eqnarray}
where the effect of the dark background is absorbed into $\widetilde{E} \equiv \sqrt{\widetilde{\bm{p}}^2 + m^2} =   E^{}_{\pm} \mp g^{}_{V} V^{}_{0}$ with $\widetilde{\bm{p}} = \bm{p} \mp g^{}_{V} \bm{V}$, such that $\widetilde{E}^2 - \widetilde{\bm{p}}^2 = m^2$. 
Here, it is equivalent to replace $\hat{\bm{p}}\cdot \bm{\Sigma}$ with $\gamma^{}_{5} \slashed{S}$, where $ \slashed{S} \equiv (|\bm{p}^{}_{}|/m^{}_{} , \widetilde{E^{}_{}} \hat{\bm{p}}^{}_{}/m^{}_{})$.
Note that we have not yet summed over the spin $s$ in the above expressions, and the standard results can be easily obtained by summing over $s$. 
These relations should be used along with the expansion
\begin{eqnarray}
% \label{eq:}
\nu(x) =   \sum^{}_{s} \int \frac{\mathrm{d}^3 \bm{p}}{(2\pi)^{3/2}} & &\sqrt{\frac{1}{2\widetilde{E}}} \left[  b^{}_{p,s} u(\bm{p},s) {e}^{- i p^{}_{+}\cdot x}  + d^\dagger_{p,s} v(\bm{p},s) {e}^{ i p^{}_{-} \cdot x} \right]  ,
\end{eqnarray}
or equivalently the form
\begin{eqnarray}
\label{eq:nux}
\nu(x) =   \sum^{}_{s} \int & & \frac{\mathrm{d}^3 \widetilde{\bm{p}}}{(2\pi)^{3/2}} \sqrt{\frac{1}{2\widetilde{E}}} \left[  b^{}_{p,s} u(\bm{p},s) {e}^{- i \tilde{p} \cdot x} + d^\dagger_{p,s} v(\bm{p},s) {e}^{ i \tilde{p} \cdot x} \right] {e}^{ -i {g^{}_{V}}V\cdot x} \;.
\end{eqnarray}
Eqs.~(\ref{eq:uuv}), (\ref{eq:uubarv}) and (\ref{eq:nux}) indicate that one may think of all the relations with $\tilde{p} = \{\widetilde{E}, \widetilde{\bm{p}}\}$ similar as those in the vacuum. The net impact of the dark background is adding an overall phase ${\rm exp}({ -i {g^{}_{V}}V^{}_{0} \, t + i {g^{}_{V}} \bm{V}\cdot \bm{x}})$ to the neutrino field. Since other fields (e.g., n, p and $e$) do not feel this phase, it will enter into the factor $\delta^4( \cdots \pm g^{}_{V} V)$ which imposes  energy momentum conservation.

Ultimately, the Hamiltonian of neutrino field is found to be
\begin{eqnarray}
% \label{eq:}
\mathcal{H} =  \int \mathrm{d}^3 \bm{p} \sum^{}_{s} \left( E^{}_{+} b^\dagger_{p,s} b^{}_{p,s} +  E^{}_{-} d^\dagger_{p,s} d^{}_{p,s} \right)  . \hspace{0.6cm} 
\end{eqnarray}

\section{\label{sec:C}Some remarks on the ground state in the dark sea}
\noindent
The formation of the background field typically takes place on cosmological time scales, say 1 Gyr corresponding to $\sim 1/(2 \times 10^{-32}~{\rm eV})$, which is significantly larger than the Compton frequency of neutrinos,  i.e., the inverse of mass $1/m^{}_{\nu} \approx 6.6 \times 10^{-15}~{\rm s}$ for $m^{}_{\nu} = 0.1~{\rm eV}$. 
The neutrino modes will therefore always stay in their energy eigenstates during the adiabatic formation of the background field, meaning that the eigenvalues of neutrino energy should change in a continuous and smooth manner without transitions.

%%%%%%%%%%%%%%%%%%%%%%%%%%%%%%%%%%%%%%%%%%%%%%%%%%%
\begin{figure}[t!]
\begin{center}
	\includegraphics[width=0.42\textwidth]{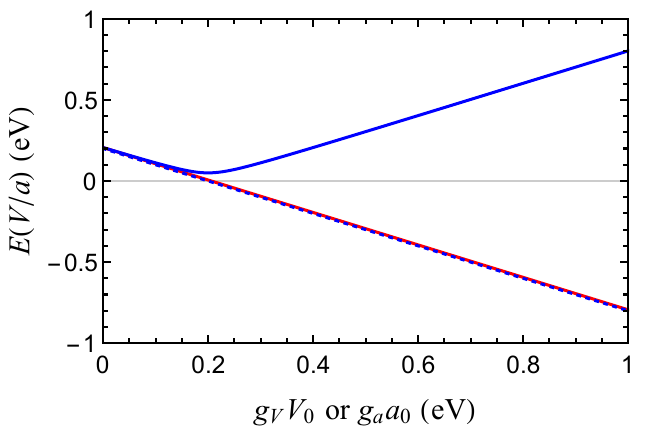}
	\includegraphics[width=0.42\textwidth]{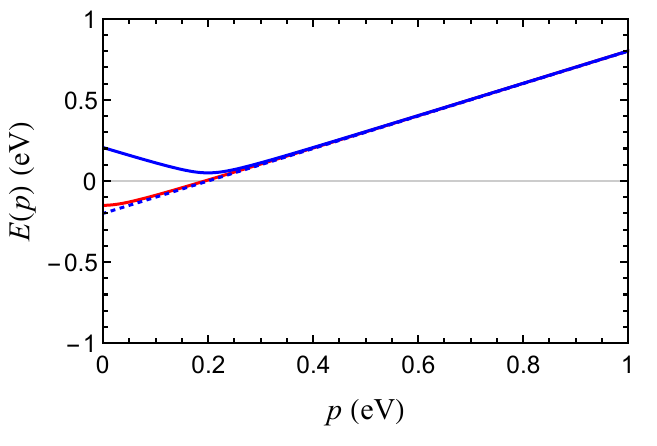}
\end{center}
\vspace{-0.3cm}
\caption{Left panel: The evolution of neutrino energy as the dark potential $g^{}_{V} V^{}_{0}$ (red curve) or $g^{}_{a} a^{}_{0}$ (blue curves) adiabatically increases. The neutrino momentum is taken as $p = 0.2~{\rm eV}$.
	Right panel: The dispersion relation with respect to $p$ with $g^{}_{V} V^{}_{0}$ (red curve) or $g^{}_{a} a^{}_{0}$ (blue curves) being $0.2~{\rm eV}$. 
	In both panels, for the solid curves the neutrino mass has been set to $m^{}_{} = 0.05~{\rm eV}$, while for the dotted blue one $m^{}_{} = 0~{\rm eV}$.}
\label{fig:EVa}
\end{figure}
%%%%%%%%%%%%%%%%%%%%%%%%%%%%%%%%%%%%%%%%%%%%%%%%%%%

For the vector and axial-vector cases, let us investigate in more detail how the energy of a neutrino mode evolves when the background field $g^{}_{V} V^{}_{0}$ or $g^{}_{a} a^{}_{0}$ gradually changes from zero to a certain value.
Their dispersion relations for the right-helicity antineutrino (corresponding to $\overline{\nu^{}_{\rm L}}$ in the massless limit) are recast as follows:
\begin{eqnarray}
\label{eq:EV}
E(V) & = & \sqrt{p^2 + m^2} - g^{}_{V}V^{}_{0} \; , \\ 
\label{eq:Ea}
E(a) & = & \sqrt{(p -  g^{}_{a}a^{}_{0} )^2 + m^2} \; ,
\end{eqnarray}
where $p \geq 0$ represents the magnitude of neutrino momentum.
As long as $m \neq 0$, these are indeed smooth functions of the potential field.
To be specific, we take the neutrino mass as $m^{}_{} = 0.05~{\rm eV}$ and set the neutrino momentum to be $p = 0.2~{\rm eV}$. Then let $g^{}_{V} V^{}_{0}$ and $g^{}_{a} a^{}_{0}$ adiabatically change from $0~{\rm eV}$ to $1~{\rm eV}$. The evolution of energy is shown in upper panel Fig.~\ref{fig:EVa}.
In the lower panel, we fix $g^{}_{V} V^{}_{0}$ and $g^{}_{a} a^{}_{0}$ as $0.2~{\rm eV}$ and vary $p$.

For comparison, in both panels of Fig.~\ref{fig:EVa} we give the case of axial-vector potential with vanishing neutrino mass as dotted curves.
Special care should be taken when the neutrino mass is vanishing, i.e., $m=0$ in Eq.~(\ref{eq:Ea}). By taking the derivative of Eq.~(\ref{eq:Ea}), we have
\begin{eqnarray}
\frac{\partial E}{\partial (g^{}_{a}a^{}_{0})} & = & -\frac{\partial E}{\partial p}= \frac{ g^{}_{a} a^{}_{0}-p}{\sqrt{(p -  g^{}_{a}a^{}_{0} )^2 + m^2}} \; . \hspace{0.5cm}
\end{eqnarray}
It is clear that as long as $m \neq 0$, the energy $E$ is a smooth function of $a^{}_{0}$ and $p$. However, when $m = 0$, the derivative becomes ill-defined at $p = g^{}_{a} a^{}_{0}$. A smooth solution to the massless case in the axial-vector potential should be
\begin{eqnarray}
E(a)  & = & p - g^{}_{a} a^{}_{0} \;,
\end{eqnarray}
which becomes identical to the vector scenario in Eq.~(\ref{eq:EV}). This is exactly what we expect when the neutrino mass is vanishing, for which the difference of results between vector and axial-vector scenarios is supposed to vanish.

%%%%%%%%%%%%%%%%%%%%%%%%%%%%%%%%%%%%%%%%%%%%%%%%%%%
\begin{figure}[t!]
\begin{center}
	\includegraphics[width=0.5\textwidth]{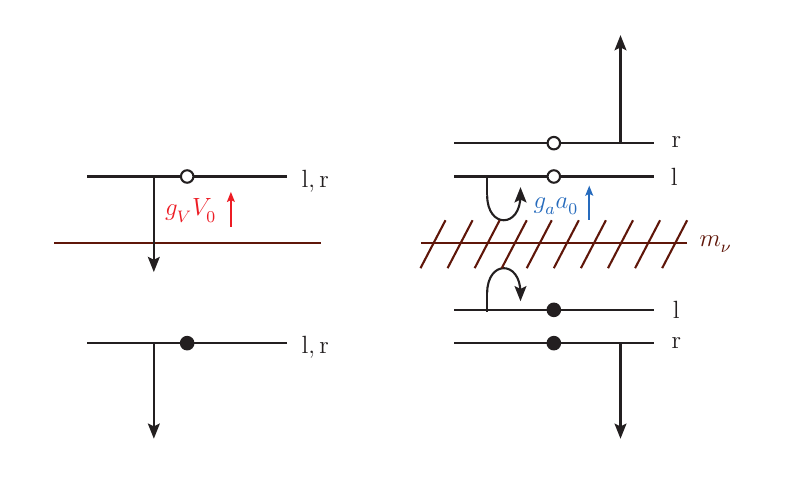}
\end{center}
\vspace{-0.3cm}
\caption{An illustration of the evolution of the neutrino energy eigenstates, as the magnitude of dark potential $g^{}_{V} V^{}_{0} < 0$ (left panel) or $g^{}_{a} a^{}_{0} < 0$ (right panel) is adiabatically increasing. The baseline of $E^{}_{\nu} = 0$ is set by the brown line. The state with an empty circle corresponds to the neutrino, while that with the filled circle to  the antineutrino. The energy is split for the left-helicity (`l') and right-helicity (`r') states in the axial-vector case. Neutrinos are not allowed to stay in the shaded region on the right.}
\label{fig:energy_level}
\end{figure}
%%%%%%%%%%%%%%%%%%%%%%%%%%%%%%%%%%%%%%%%%%%%%%%%%%%

The adiabatic evolution of the eigenstates of neutrinos is schematically shown in Fig.~\ref{fig:energy_level} for the vector and axial-vector cases, respectively. We explain the figure quantitatively in what follows and note that we have verified it by numerically solving the Dirac equation. 
For the axial vector case, there is an energy barrier set by the neutrino mass which keeps the neutrino state above the zero-point energy, as $g^{}_{a} a^{}_{0}$ adiabatically increases.  
In the massless limit, such a barrier does not exist, and the left-handed field shifts smoothly down as in the vector case. Because only the left-handed neutrino field is responsible for the beta decays (left-helicity for neutrino and right-helicity for antineutrino in Fig.~\ref{fig:energy_level}), the effects of axial-vector and vector dark potentials on the beta-decay spectrum should be the same for $m^{}_{\nu} = 0$.

For the axial-vector potential, the adiabatic approximation will break down
in the extremely narrow parameter space $0< m^{}_{\nu} \lesssim 10^{-32}~{\rm eV}$ (for which the background field changes faster than the neutrino mass), and one may expect the probability of tunneling crossing the mass barrier (for the massless case without barrier, the tunneling probability is equivalently one). 
This is very similar to the matter effect of neutrino oscillations in varying matter profile~\cite{Kuo:1989qe}. On the resonance, when the adiabaticity parameter is large $\gamma \gg \mathcal{O}(1)$, the neutrino will always stay in one specific mass eigenstate. But for $\gamma \lesssim \mathcal{O}(1)$, transition occurs from one neutrino mass eigenstate to another.
However, we should notice that at least two neutrino mass eigenvalues should be larger than $0.0086~{\rm eV}$ according to neutrino oscillation data, and hence the transition for them is always adiabatic.
The above discussion is only relevant for cosmological time scales. For the time scale relevant to KATRIN runs, the dark potential explored here does not change.

In Fig.~1 of the main text, in the limit of $m^{}_{\nu} \to 0$, the result of the axial-vector case seems unable to continuously transit to that of the vector one.
This is not a surprise considering that the cosmological time scale, one billion years corresponding to $2 \times 10^{-32}~{\rm eV}$, separates the massless and sizable massive cases.

%KATRIN has achieved an excellent concentration of ${\rm T}^{}_{2}$ over other isotopes. The concentration of ${\rm T}^{}_{2}$ for the second campaign (first campaign) is $97.3\%$ ($95.3\%$), while those of the isotopes ${\rm D T}$ and ${\rm H T}$ have been purified to extremely low values $0.31\%$ ($1.1\%$) and $2.3\%$ ($3.5\%$), respectively. Such amount of contamination should be negligible in the current published results of KATRIN, e.g., see Fig.~4 of Ref.~\cite{Doss:2006zv}. Hence 

%For the scalar and pseudoscalar cases, to resolve the degeneracy between the new physics effect and the vacuum mass, we will adopt the Planck limit on the sum of neutrino masses in vacuum~\cite{Planck:2018vyg}. The Planck chisquare is available by transforming from the likelihood via $\chi^2 = -2 \ln{L}$, which is well approximated by
%\begin{eqnarray}
% \label{eq:}
%\hspace{-0.3cm}\chi^2_{\rm Planck}(\Sigma) = 200 \left(\frac{\Sigma}{\rm eV} \right)^2 + 12.4 \left(\frac{\Sigma}{\rm eV} \right), 
%\end{eqnarray}
%corresponding to the upper limit $\Sigma < 0.12~{\rm eV}$. This strong limit actually forces the vacuum neutrino mass to be minimal. Combining $\chi^2_{\beta}$ and $\chi^2_{\rm Planck}(\Sigma)$, we generate the exclusion for the scalar and pseudoscalar potentials. We checked that the results are almost the same as the case by just taking  $m^2_{\nu}$ to be vanishing.

\bibliographystyle{utcaps_mod}

\bibliography{reference}

\end{document}